\documentclass[prl, twocolumn, 10pt, aps, superscriptaddress, floatfix, showpacs, citeautoscript]{revtex4-1}

\usepackage{graphicx}
\usepackage{amsmath}
\usepackage[usenames, dvipsnames]{color}
\usepackage[normalem]{ulem}

\newcommand{\InAsGaSb}{$\mathrm{InAs}/\mathrm{GaSb}$ }
\newcommand{\HgTe}{$\mathrm{HgTe}$ }
\newcommand{\HgTeCdTe}{$\mathrm{HgTe}/\mathrm{CdTe}$ }
\newcommand{\kp}{$\mathbf{k}{\cdot}\mathbf{p}$ }

\begin{document}

\title{Robust Helical Edge Transport in Quantum Spin Hall Quantum Wells}

\author{Rafal Skolasinski}
\affiliation{QuTech and Kavli institute of nanoscience, Delft University of Technology, 2600 GA Delft, The Netherlands}

\author{Dmitry I. Pikulin}
\affiliation{Station Q, Microsoft Research, Santa Barbara, California 93106-6105, USA}

\author{Jason Alicea}
\affiliation{Department of Physics and Institute for Quantum Information and Matter,
California Institute of Technology, Pasadena, CA 91125, USA}
\affiliation{Walter Burke Institute for Theoretical Physics, California Institute of Technology, Pasadena, CA 91125, USA}

\author{Michael Wimmer}
\affiliation{QuTech, Delft University of Technology, 2600 GA Delft, The Netherlands}

\date{\today}

\begin{abstract}
We show that burying of the Dirac point in semiconductor-based quantum-spin-Hall systems can generate unexpected robustness of edge states to magnetic fields. A detailed ${\bf k\cdot p}$ band-structure analysis reveals that \InAsGaSb and \HgTeCdTe quantum wells exhibit such buried Dirac points.
By simulating transport in a disordered system described within an effective model, we further demonstrate that buried Dirac points yield nearly quantized edge conduction out to large magnetic fields, consistent with recent experiments.
\end{abstract}

\maketitle

{\bf \emph{Introduction.}}~Topological insulators (TIs) are materials that exhibit a gapped bulk yet enjoy metallic surface or edge states protected by time-reversal symmetry.
In particular, two-dimensional (2D) TIs host helical edge modes---i.e., counter-propagating states composed of Kramers partners---that underlie quantized edge conductance~\cite{kane_quantum_2005, Hasan2010, Qi2011}.
Consequently, 2D TIs are often referred to as quantum spin Hall (QSH) systems.
The experimentally most studied QSH systems are now based on semiconductor quantum wells.
Following the proposal of Bernevig, Hughes, and Zhang~\cite{bernevig_quantum_2006}, the QSH effect was first observed in HgTe/(Hg,Cd)Te quantum wells~\cite{konig_quantum_2007}; various QSH signatures, including quantized edge transport, have by now been identified in this material \cite{Roth2009, Brune2012, Nowack2013, Hart2014}.

In HgTe, the QSH effect originates from an inversion of electron and hole bands that is intrinsic to HgTe.
This inversion can also be engineered in a multilayer quantum well.
In particular, \InAsGaSb quantum wells were also predicted to be QSH systems~\cite{liu_quantum_2008}, as they exhibit a so-called broken gap alignment where the conduction band edge of electrons is energetically below the valence band edge of holes.
Quantized edge conductance has also been observed in \InAsGaSb\cite {knez_evidence_2011, knez_observation_2014, du_robust_2015}, and the properties of the band inversion and edge-state transport have since been investigated by several experimental groups~\cite{Pribiag2015,Qu2015, Mueller2015, Couedo2016, Karalic2016, Du2017, Tiemann2017}.

The hallmark quantized edge conductance in QSH systems originates from time-reversal symmetry, which prevents the helical edge states from elastically backscattering in the presence of non-magnetic disorder.  A magnetic field $B$ breaks time-reversal symmetry, and common expectation dictates that quantized conductance must break down in this case.
For example, a magnetic field applied to semiconductor-based QSH systems can directly couple the counter-propagating edge modes, opening up a Zeeman gap in the edge spectrum.
It thus came as a surprise that Ref.~\cite{du_robust_2015} measured edge conductances that remained quantized with in-plane magnetic fields up to $12\,$T---sharply defying theoretical expectations.

\begin{figure}
\includegraphics[width=0.9\linewidth]{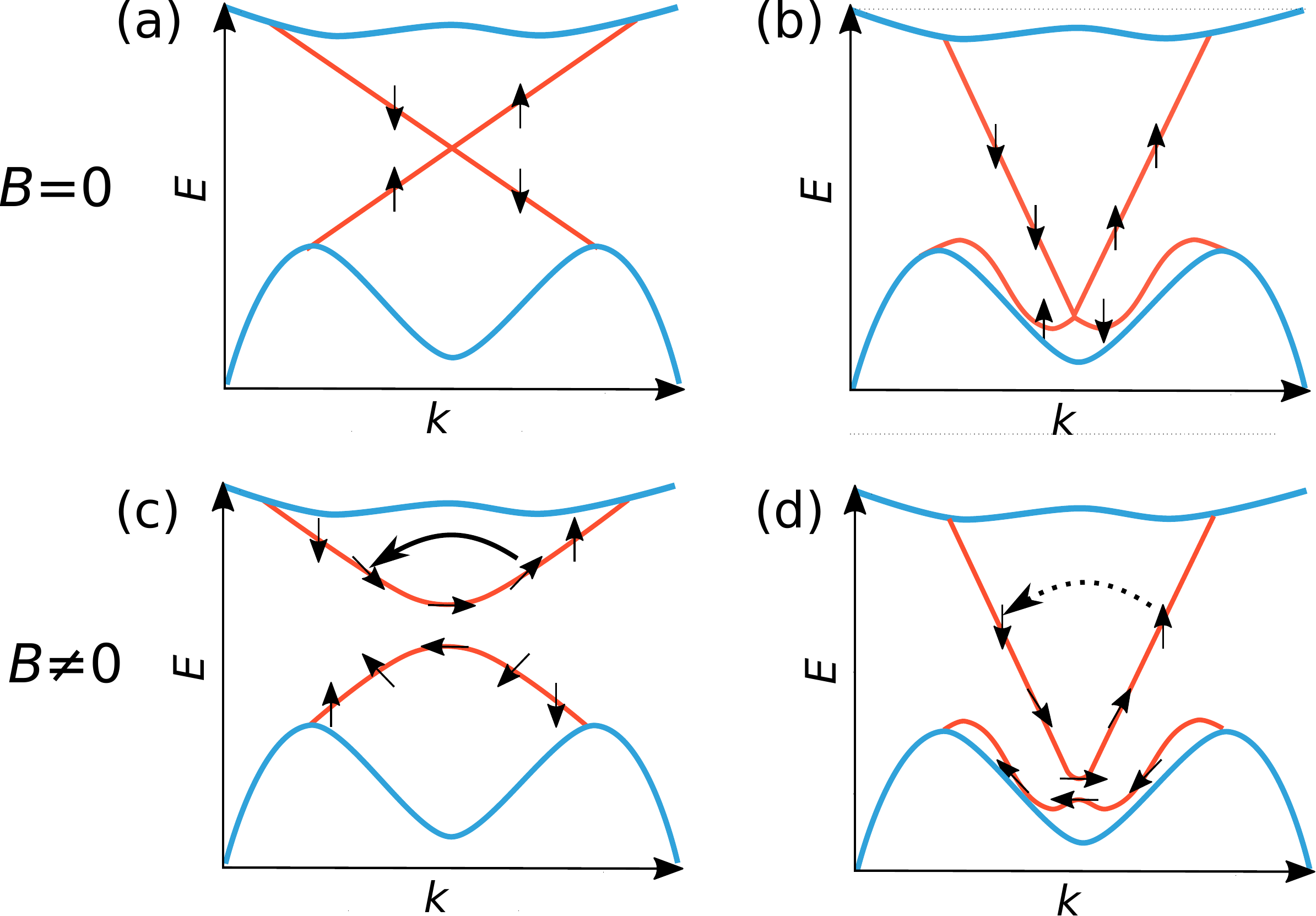}
\caption{
  Schematic depiction of edge-state dispersions: in the absence of a   magnetic field, the edge-state crossing is topologically protected, but may (a) reside in the gap or (b) be hidden in a bulk band.
  In a finite magnetic field, a Zeeman gap opens and the edge-state spins become canted---permitting backscattering as in (c)
  However, when the edge-state crossing is hidden in a bulk band, spins within the gap are further away from the Zeeman gap and nearly anti-align, greatly suppressing backscattering as in (d).
}
\label{fig:scheme}
\end{figure}

Here we show that, contrary to naive expectations, edge-state transport in semiconductor-based QSH systems (\HgTe and $\mathrm{InAs}/\mathrm{GaSb}$) \emph{typically exhibits a very weak dependence on in-plane magnetic fields.}
We have identified three mechanisms for such robustness: (i) The effective edge-state $g$-factor is strongly suppressed compared to the bulk electron $g$-factor due to significant heavy-hole contribution in the edge-state wavefunction.
(ii) The Dirac point of the edge states typically resides not in the bulk energy gap, but is hidden in a bulk band.
A Zeeman gap opened by the magnetic field appears only at the Dirac point and is thus invisible to transport (see Fig.~\ref{fig:scheme}).
(iii) Although the combination of disorder and a magnetic field generically permits backscattering, it is strongly suppressed away from the Dirac point due to the nearly anti-aligned spins of the counter-propagating edge states
[see Figs.~\ref{fig:scheme}(b) and (d)].
This alignment increases for energies away from the Dirac point.
When the Dirac point is buried, one then obtains near-perfect quantization of edge conductance in a disordered system out to large magnetic fields of order $10\,$T as observed experimentally.

We note that buried Dirac points have been predicted and observed in several three-dimensional TIs~\cite{Zhang2009, Hsieh2009, Brune2011}.
Our findings suggest that Dirac-point burial is a common feature also in 2D QSH quantum-well platforms.

{\bf \emph{Suppression of $g$-factor.}}~We first flesh out the suppression of the edge-state $g$-factor, which is already accessible from the canonical Bernevig-Hughes-Zhang (BHZ) model~\cite{bernevig_quantum_2006} written as:
\begin{align}
[M - B_+ (k_x^2 - \partial_y^2)]\psi_1 + A (k_x - \partial_y)\psi_2 = E \psi_1;\label{eq:BHZ1}\\
A (k_x + \partial_y)\psi_1 - [M - B_- (k_x^2 - \partial_y^2)]\psi_2 = E \psi_2. \label{eq:BHZ2}
\end{align}
Here $A, M$, and $B_\pm = B\pm D$ are BHZ model parameters, $x$ is the propagation direction, $y$ is the direction into the QSH bulk, and $\psi_{1,2}$ respectively denote the electron and hole part of the wavefunction within one spin sector.
The derivation of the effective $g$-factor is based on computing the wavefunctions $\psi_{1,2}$ with a hard-wall boundary condition, similar to Ref.~\cite{Zhou2008}, and is presented in the Appendix.
The result is simple and is based on the relative contributions of electrons and holes in the edge wavefunctions:
\begin{align}
g_{\rm{eff}} = \frac{g_e B_- + g_h B_+}{B_+ + B_-}.
\label{eq: edge-g-factor}
\end{align}
Here $g_e$ and $g_h$ are electron and hole $g$-factors, respectively.
Equation \eqref{eq: edge-g-factor} shows that the effective $g$-factor of the edge states is the weighted sum of the electron and hole $g$-factors with their corresponding inverse masses as pre-factors.

Typically, $g_h$ is much smaller than $g_e$;
in fact $g_h = 0$ by symmetry in [001] quantum wells~\cite{Winkler}.
Moreover, the hole mass usually far exceeds that of electrons, i.e., $B_- \ll B_+$.
Together these properties suppress the effective edge-state $g$-factor considerably compared to bulk values.

We have performed \kp simulations (see Appendix for details) to obtain numerical values for the $g$-factor in experimentally relevant geometries.
For \InAsGaSb we find an edge-state $g$-factor $g_\text{eff}\sim 2$, whereas for \HgTe we find $g_\text{eff}\sim 8-10$ (in our conventions the Zeeman gap is $g_\text{eff} \mu_B B$, with $\mu_B$ the Bohr magneton).
In contrast, the bulk electron $g$-factors are $g\sim 6-8$ in \InAsGaSb and $g\sim 30-60$ in $\mathrm{HgTe}$.

{\bf \emph{Dirac-point burial from \kp models.}}~In the `pure' BHZ model given above, the edge-state Dirac point
always resides in the gap~\cite{Zhou2008}.
Recovering the burial of the Dirac point requires going beyond this minimal model.
To this end we now simulate the full semiconductor heterostructure for the experimentally relevant \InAsGaSb and \HgTeCdTe quantum wells.
In the numerical analysis we use the $8\times 8$ Kane Hamiltonian~\cite{kane_band_1957, kane_semiconductors_1966, bastard_wave_1988}.
Details of the model and material parameters appear in the Appendix.
Using a finite-difference method with grid spacing $a$, we convert the continuous Kane Hamiltonian into a tight-binding model.
The resulting energy dispersion are then computed using Kwant~\cite{kwant}.

\begin{figure}
\includegraphics[width=\columnwidth]{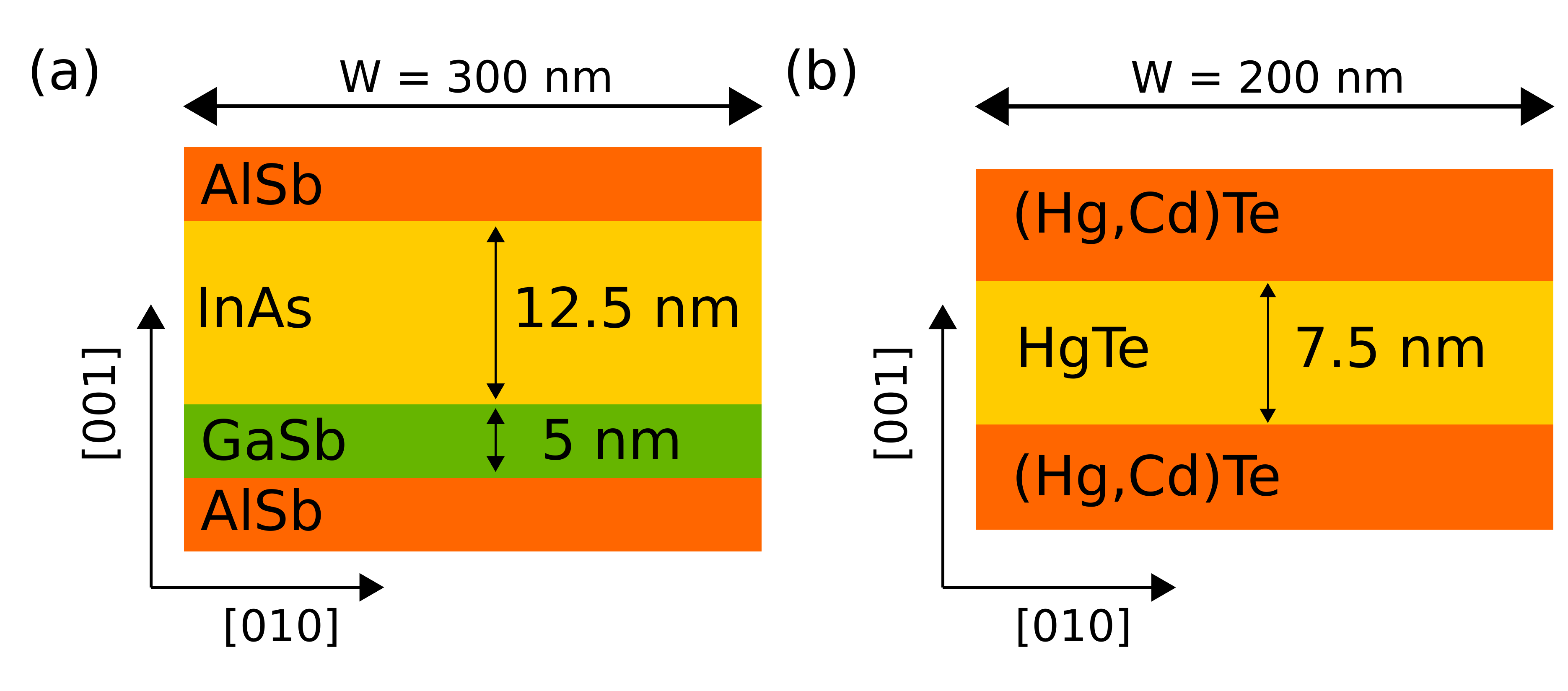}
\includegraphics[width=\columnwidth]{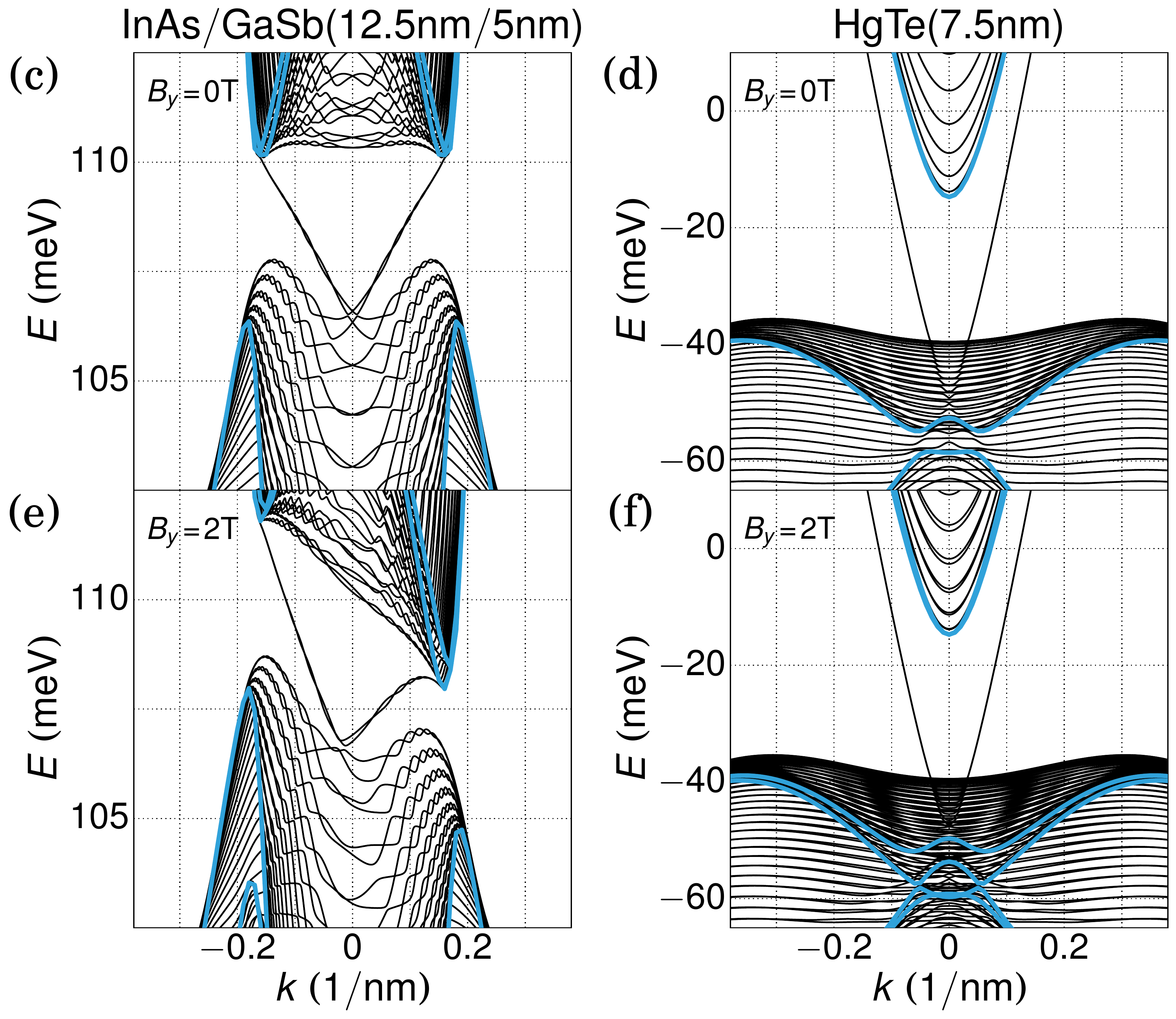}
\caption{
(a,b) system geometries used for \kp simulations.
(c-f) Band structures for \InAsGaSb (c,e) and HgTe/CdTe (d,f).
For both materials we observe Dirac points buried in a valence
band, which obscures the opening of a Zeeman gap under applied
in-plane magnetic fields as in (e) and (f).
}
\label{fig:kp_disps}
\end{figure}

We investigate [001]-grown quantum wells sketched in Figs.~\ref{fig:kp_disps}(a) and (b).
In particular, we consider \InAsGaSb with AlSb barrier (layer thicknesses $12.5\,$nm/$5\,$nm as in Ref.~\cite{Qu2015}), and \HgTe with HgCdTe barriers (thickness $7.5\,$nm as in Refs.~\cite{Roth2009, Hart2014}).
Figure~\ref{fig:kp_disps} shows the dispersion for these heterostructures along the [100] direction.  
We compare the dispersion for an infinite 2D quantum well without edges
(blue lines) to systems of finite width $W$ (black lines) modeled using hard-wall boundary conditions.

Figures~\ref{fig:kp_disps}(c) and (d) respectively illustrate the energy dispersions for \InAsGaSb and \HgTe in the absence of a magnetic field.
In both quantum wells we observe that the edge-state crossing is shifted out of the topological gap and buried in the valence band.
Note that while the crossing itself is topologically protected, its position inside the gap is not.

The \kp results diverge from the BHZ model due to the presence of additional hole states that are close in energy to the electron and heavy-hole (HH) bands forming the inverted band structure.
For $\mathrm{InAs}/\mathrm{GaSb}$, those states lead to a significant deviation of the band structure at the topological gap from the BHZ model, which only contains momentum up to second order.
Those states are energetically much further away from the gap than the size of the gap itself [no additional hole states are visible in Fig.~\ref{fig:kp_disps}(c)]; nevertheless, they strongly influence the gap edges at finite momentum, as the coupling between bands increases with momentum (see Appendix
for models that take into account this interaction).
In the case of \HgTe a second HH band crosses with the topological gap.
Since it only weakly interacts with the edge state, the Dirac point is deeply hidden in this additional band.

Figures~\ref{fig:kp_disps}(e) and (f) show the energy dispersions in a finite
magnetic field.
For both quantum wells, the Zeeman splitting of the edge states remains well-hidden in the valence band.
Note that while the \InAsGaSb bulk band structure and bulk transport therein is affected by an in-plane field due to orbital effects on the tunneling between the two layers~\cite{Yang1997, Qu2015}, this modification neither removes the edge states~\cite{Hu2016} nor the position of the edge-state crossing \footnote{One can see from Fig.~\ref{fig:kp_disps}(e) that the parallel field generates indirect gapless bulk excitations.  These bulk states are, however, expected to be more susceptible to localization compared to the edge states.}.

\begin{figure}
\includegraphics[width=\linewidth]{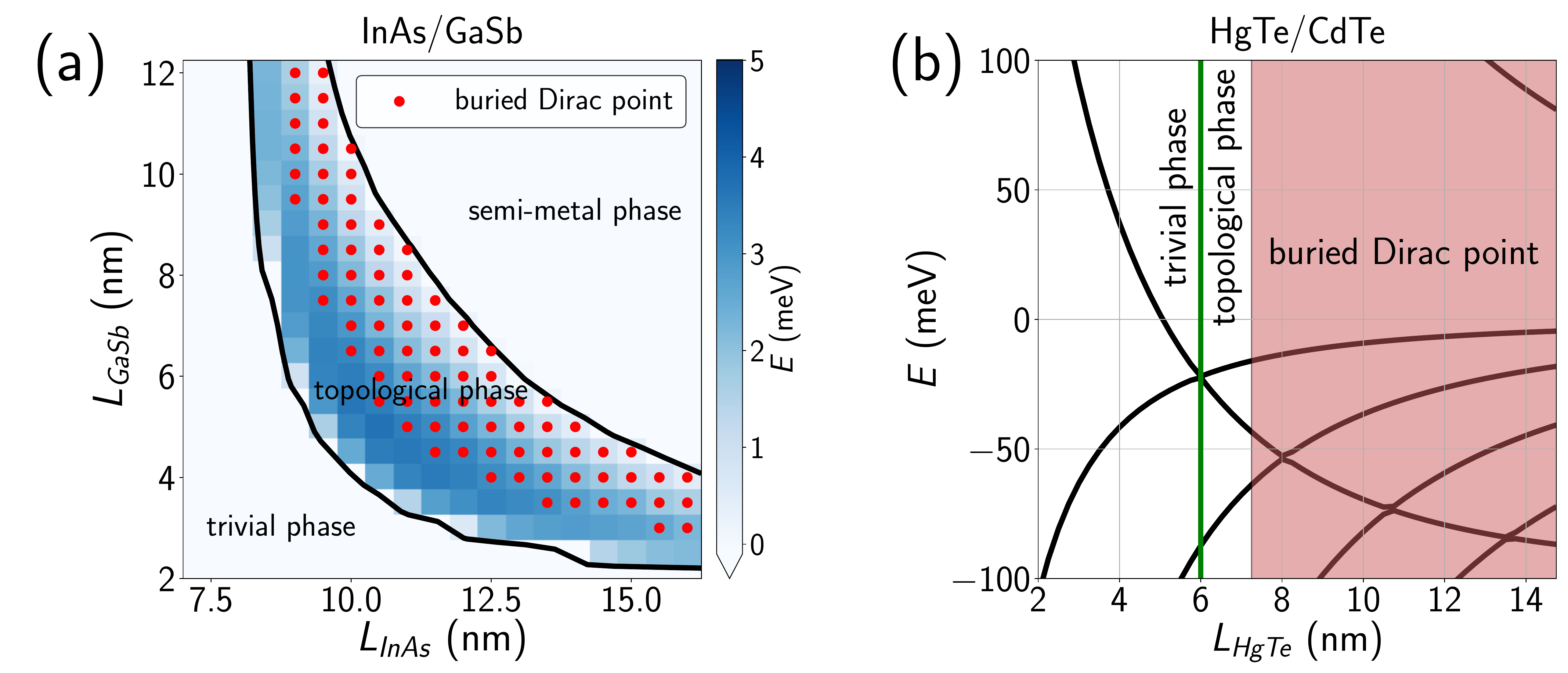}
\caption{
  (a) Topological gap of \InAsGaSb as a
  function of InAs and GaSb well thicknesses.
  A red dot indicates a
  buried Dirac point.
  (b) Subband edges at the $\Gamma$-point of \HgTe
  as a function of \HgTe well thickness.
  The Dirac point is buried for
  thickness $L_{HgTe}\geq7.25\,$nm.}
  \label{fig:phasediagrams}
\end{figure}

Figure~\ref{fig:phasediagrams} summarizes our simulations for different quantum-well thicknesses: Fig.~\ref{fig:phasediagrams}(a) shows the topological phase diagram of \InAsGaSb as a function of layer thicknesses (a non-monotonic behavior of the topological gap was also previously found in Ref.~\cite{Hu2016a}), while Fig.~\ref{fig:phasediagrams}(b) shows the HgTe band edges as a function of layer thickness (here we only have one parameter).
In both cases we indicate when the edge-state Dirac point is buried---which occurs for most of the topological phase space as expected from our general arguments.
The edge-state crossing remains in the gap only close to the topological phase transition; here only two bands interact in a small range of momentum and can be well-described by the BHZ model.

{\bf \emph{Modeling Dirac-point burial via edge potentials.}}~So far we have considered the edge of the 2D QSH systems simply as a hard wall.
However, several semiconducting surfaces additionally exhibit a band bending at the interface.
A prominent example is InAs where the band bending can be of the order of $100\,$meV~\cite{Tsui1970, Noguchi1991}.
In fact, band bending has been shown to have significant effects also in \InAsGaSb quantum wells~\cite{Nichele2016, Mueller2017}.
Apart from band bending due to details of the semiconductor surface, gating can also lead to a non-uniform electrostatic potential near the surface,
e.g.,~due to the change of dielectric constant at the semiconductor/vacuum interface.

A position-dependent potential $V(y)$ that changes only close to the surface (edge potential) has a strong effect on the edge-state dispersion: within first-order perturbation theory it leads to a shift $\Delta E(k_x)= \left<\psi(k_x)\right|V\left|\psi(k_x)\right>$.
In particular, since bulk states are affected little by the edge potential, the edge-state crossing is shifted by $\Delta E(k_x=0)$ with respect to the bulk bands.
Thus, if the edge potential is much larger than the topological gap, it also leads to a burying of the Dirac point.
(The edge potential may also give rise to trivial edge states that are also expected to be insensitive to a magnetic field.
In contrast to topological edge states these are not expected to be protected from scattering, leading to a length dependence of the edge conductance~\cite{Nichele2016}.)

\begin{figure}
\includegraphics[width=\columnwidth]{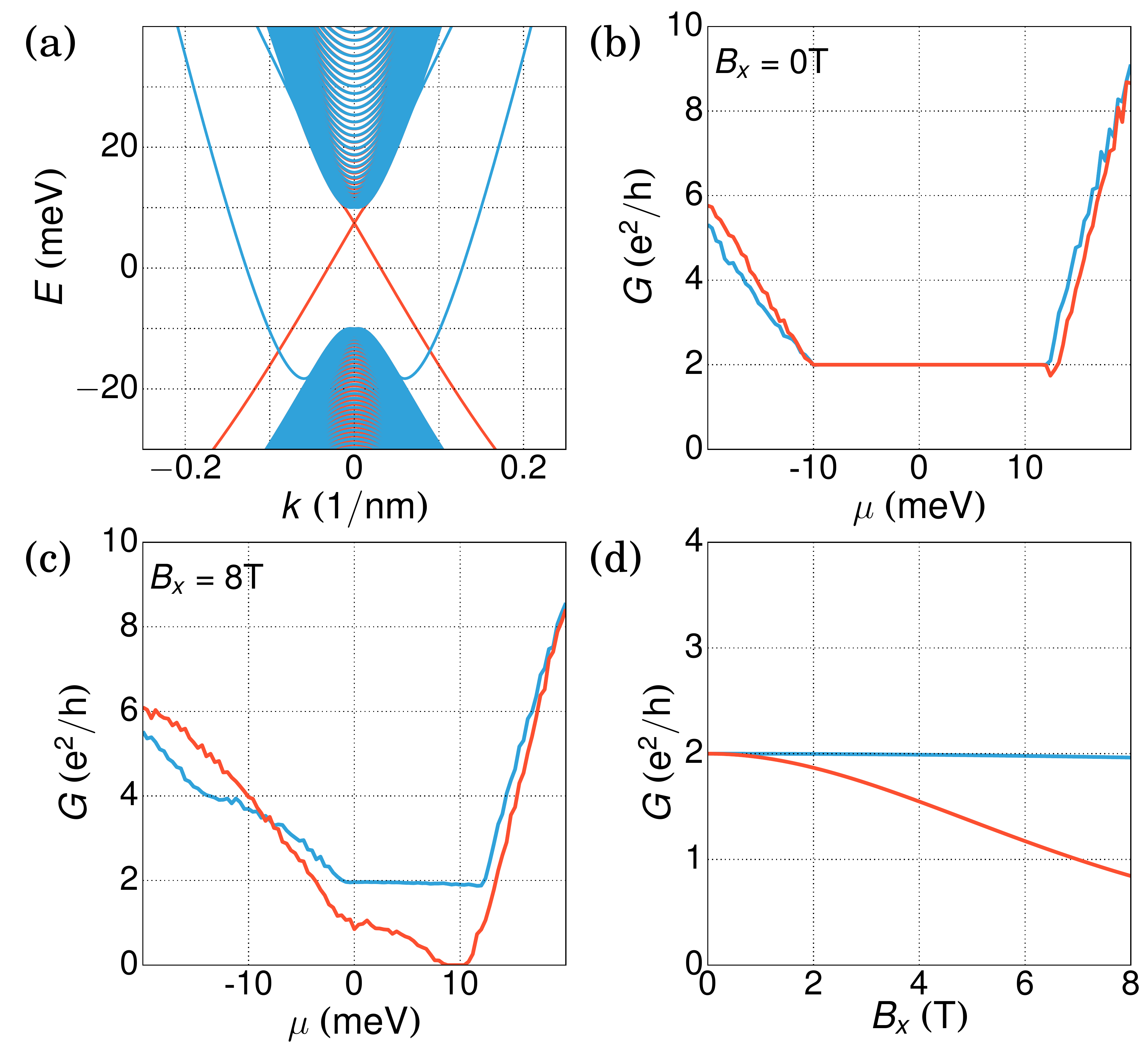}
\caption{
(a) Band structure and (b-d) transport calculations for the BHZ model with (blue) and without (red) an additional edge potential $V_\text{edge}$.
Transport calculations were performed for a disordered system at (b) zero field and (c) with an in-plane field $B_x=8\,$T.
(d) Transport calculation at fixed $\mu=3~\mathrm{meV}$.
All transport calculations are averaged over 50 different disorder realizations, with parameters $V_\text{edge}=-0.14\,$eV, $U_0=0.05\,$eV, $L=4000\,$nm, $W=1000\,$nm, and finite difference grid spacing $a=4\,$nm.
}
\label{fig:transport}
\end{figure}

Figure~\ref{fig:transport} shows the burying of the edge-state Dirac point obtained from a finite-width BHZ model supplemented by an edge potential.
We use the BHZ parameters for HgTe of Ref.~\cite{Konig2008} and a finite-difference tight-binding model, with an extra potential $V_\text{edge}$ at the outermost lattice point.
For $V_\text{edge}=0$ (red lines) we find the usual dispersion with the edge-state crossing in the band gap.
A finite $V_\text{edge}\neq 0$ (blue lines) leaves the bulk states nearly unchanged, but indeed moves the edge-state crossing into the bulk.

Apart from potentially being physically present in semiconductor devices, we can also use the edge potential purely as a tool that leads to a Dirac-point burial within the BHZ model.
This is particularly advantageous for numerical calculations, which are far more costly for a 3D \kp model.

{\bf \emph{Quantized conductance in strong in-plane magnetic fields.}}~So far we have emphasized generic mechanisms for hiding the edge-state Zeeman gap within a bulk band.
In such cases, observing a clear field-induced edge-state gap through transport would certainly be challenging.
Yet, time-reversal symmetry is broken by an in-plane magnetic field, and backscattering from disorder is allowed also \emph{outside} the edge-state Zeeman gap.
Naively, a magnetic field should thus lead to an appreciable breakdown of the conductance quantization.

We will now argue that, in practice, conductance may stay nearly quantized even in very strong magnetic fields ($B\geq 10$T): from Fermi's golden rule we find that the mean free path of edge states in a disordered potential is given as~\cite{Pikulin2014}
\begin{equation}\label{eq:mfp}
l_\text{tr} = \frac{c \hbar^2 v_\text{F}}{V_\text{dis}^2\xi} \left(\frac{\delta\mu}{g_\text{eff} B}\right)^2\,.
\end{equation}
Here, $v_\text{F}$ is the edge-state velocity, $c$ is a numerical factor $\sim 1$,
$\delta\mu$ is the energy with respect to the edge-state crossing, and we assumed uncorrelated disorder $\left<V(x) V(x')\right> =V_\text{dis}^2 \xi \delta(x-x')$.
In the bulk-insulating regime, burial of the edge-state crossing implies that $\delta\mu$ must be of order or larger than the gap size.
Together with the strong suppression of the edge-state $g$-factor $g_\text{eff}$ discussed earlier, $(\delta\mu/g_\text{eff} B)^2$ then is a large factor.
Physically, this suppression of backscattering away from Dirac point originates from the fact that kinetic energy efficiently anti-aligns spins of the edge state away from the Zeeman gap even in presence of magnetic field; recall Fig.~\ref{fig:scheme}(d).
In practice, the suppression of scattering presented here may rival that
arising from bona fide topological protection at zero magnetic field.

To further quantify the suppression of backscattering, we have performed conductance calculations for a disordered BHZ model, with and without edge potential, i.e., with and without burying of the Dirac point.
As for the results sketched in Fig.~\ref{fig:transport}(a), we use the HgTe parameters from Ref.~\cite{Konig2008}, and compute transport through a rectangular region of length $L$ and width $W$.
We use a random disorder potential drawn independently for every lattice point from the uniform distribution $[-U_0/2,U_0/2]$, and compute the conductance using Kwant~\cite{kwant}.
At zero magnetic field [Fig.~\ref{fig:transport}(b)] both models show almost identical transport properties.
In particular, the conductance in the gap is perfectly quantized due to topological protection.
This behavior changes drastically once a strong in-plane magnetic field is applied [Fig.~\ref{fig:transport}(c)]: Without edge potential, the conductance drops well below the quantized value of $2e^2/h$.
Disorder leads to backscattering within the complete range of energies in the topological gap (not only the small Zeeman gap opened in the edge-state spectrum).
When the edge-state crossing is buried, by contrast, conductance inside the gap stays almost perfectly quantized.
This stark contrast can also be seen in Fig.~\ref{fig:transport}(d) where we plot conductance as a function of magnetic field for a fixed chemical potential residing in the bulk gap.

{\bf \emph{Conclusions.}}~In contrast to common expectation, we have shown that the edge-state conductance quantization in semiconductor QSH systems can be surprisingly robust against in-plane magnetic fields.
This may be a possible explanation for the surprising findings of Ref.~\cite{du_robust_2015}, and we could expect to find similar robustness in HgTe.
Our findings also highlight a challenge for proposals to use QSH edges as a Majorana platform~\cite{Fu2009, Alicea2012}: Localizing Majorana zero modes requires the ability to align the chemical potential within the edge-state Zeeman gap, which could require exceedingly large fields if the Dirac point is buried in a bulk band.
A good strategy to overcome this obstacle is to operate in a regime closer to the topological phase transition where the edge-state crossing remains in the gap (if edge potentials are unimportant).
Alternative, a side-gate might be used to apply an electrostatic potential to move the Dirac point back in the topological gap.
These strategies may also allow one to finally observe a strong in-plane magnetic field dependence that would distinguish topological from trivial edge states---the latter naturally exhibiting little field dependence.

While finishing this work, we became aware of a related preprint~\cite{Li2017} that found a hidden Dirac point in the band structure of InAs/GaSb within an effective 6-band model, in qualitative agreement with our full \kp calculations.

\begin{acknowledgments}
We acknowledge useful discussions with L.~Molenkamp, A.R.~Akhmerov and T.~Hyart. R.S.~and M.W.~were supported by the Dutch national science organization NWO. D.I.P.~acknowledges support by Microsoft Corporation Station Q. J.A.~gratefully acknowledges support from the National Science Foundation through grant DMR-1723367; the Army Research Office under Grant Award W911NF-17-1-0323; the Caltech Institute for Quantum Information and Matter, an NSF Physics Frontiers Center with support of the Gordon and Betty Moore Foundation through Grant GBMF1250; and the Walter Burke Institute for Theoretical Physics at Caltech.
\end{acknowledgments}

\bibliography{bib}

\clearpage
\section{Supplemental Material}

\subsection{\kp simulations}
We use the standard $8\times 8$ Kane Hamiltonian~\cite{kane_band_1957, kane_semiconductors_1966, bastard_wave_1988} for semiconductors in our numerical band structure calculations.
The material parameters in this Hamiltonian are position-dependent due to the layered structure, and care has to be taken to symmetrize the Hamiltonian.
Following the approach put forward by Burt and Foreman~\cite{burt_justification_1992, foreman_elimination_1997}, the Hamiltonian for the [001] growth direction reads form~\cite{novik_band_2005, pfeuffer-jeschke_transport_2000}:

\begingroup
\renewcommand*{\arraystretch}{2.5}
\begin{widetext}
\begin{equation}
\label{eq:H8x8}
H=
\begin{pmatrix}
T & 0 & -\frac{1}{\sqrt{2}} P k_{+} & \sqrt{\frac{2}{3}} P k_{z} & \frac{1}{\sqrt{6}}P k_{-} & 0 & -\frac{1}{\sqrt{3}} P k_{z} & -\frac{1}{\sqrt{3}} P k_{-} \\
0 & T & 0 & -\frac{1}{\sqrt{6}}P k_{+} & \sqrt{\frac{2}{3}} P k_{z} & \frac{1}{\sqrt{2}} P k_{-} & -\frac{1}{\sqrt{3}} P k_{+} & \frac{1}{\sqrt{3}} P k_{z} \\
-\frac{1}{\sqrt{2}} k_{-} P & 0 & U+V & -\overline{S}_{-} & R & 0 & \frac{1}{\sqrt{2}}\overline{S}_{-} & -\sqrt{2} R \\
\sqrt{\frac{2}{3}}k_z P & -\frac{1}{\sqrt{6}} k_{-} P & -\overline{S}^{\dagger}_{-} & U-V & C & R & \sqrt{2} V & -\sqrt{\frac{3}{2}} \widetilde{S}_{-} \\
\frac{1}{\sqrt{6}}k_{+} P & \sqrt{\frac{2}{3}} k_z P & R^{\dagger} & C^{\dagger} & U-V & \overline{S}^{\dagger}_{+} & -\sqrt{\frac{3}{2}}\widetilde{S}_{+} & -\sqrt{2} V \\
0 & \frac{1}{\sqrt{2}} k_{+}P & 0 & R^{\dagger} & \overline{S}_{+} & U+V & \sqrt{2}R^{\dagger} & \frac{1}{\sqrt{2}}\overline{S}_{+} \\
-\frac{1}{\sqrt{3}} k_{z} P & -\frac{1}{\sqrt{3}} k_{-} P & \frac{1}{\sqrt{2}} \overline{S}^{\dagger}_{-} & \sqrt{2} V & -\sqrt{\frac{3}{2}} \widetilde{S}^{\dagger}_{+} & \sqrt{2} R & U - \Delta & C \\
-\frac{1}{\sqrt{3}} k_{+} P & \frac{1}{\sqrt{3}} k_{z} P & -\sqrt{2} R^{\dagger} & -\sqrt{\frac{3}{2}} \widetilde{S}^{\dagger}_{-} & -\sqrt{2} V & \frac{1}{\sqrt{2}} \overline{S}^{\dagger}_{+} & C^{\dagger} & U-\Delta
\end{pmatrix},
\end{equation}
\end{widetext}

where

\begin{equation*}
k_\parallel^2 = k_x^2 + k_y^2, \quad k_\pm = k_x \pm i k_y, \quad k_z = -i \partial / \partial z,
\end{equation*}
\begin{align*}
T &= E_c + \frac{\hbar^2}{2m_0} \left(\gamma^{\prime}_0k_\parallel^2 + k_z \gamma^{\prime}_0 k_z\right), \\
U &= E_v -\frac{\hbar^2}{2m_0} \left(\gamma^{\prime}_1 k_\parallel^2 + k_z \gamma^{\prime}_1 k_z \right), \\
V &= -\frac{\hbar^2}{2m_0} \left( \gamma^{\prime}_2 k_\parallel^2 - 2 k_z \gamma^{\prime}_2 k_z \right), \\
R &= -\frac{\hbar^2}{2m_0} \frac{\sqrt{3}}{2} \left[(\gamma^{\prime}_3-\gamma^{\prime}_2)k_{+}^2 - (\gamma^{\prime}_3+\gamma^{\prime}_2)k_{-}^2\right] ,\\
\overline{S}_{\pm} &= -\frac{\hbar^2}{2m_0} \sqrt{3} k_\pm \left(\{\gamma^{\prime}_3, k_z\} + [\kappa^{\prime}, k_z]\right),\\
\widetilde{S}_{\pm} &= -\frac{\hbar^2}{2m_0} \sqrt{3} k_\pm \left(\{\gamma^{\prime}_3, k_z\} - \frac{1}{3} [\kappa^{\prime}, k_z]\right), \\
C &= \frac{\hbar^2}{m_0}k_{-} \left[\kappa^{\prime}, k_z\right].
\end{align*}
\endgroup
Here, $P$ is the Kane interband momentum matrix element, $E_c$ and $E_v$ are the conduction and valence band edges, respectively, and $\Delta$ is the spin-orbit splitting energy.
$[A, B] = AB - BA$ is the commutator and $\{A, B\} = AB + BA$ is the anticommutator for the operators A and B.

$\gamma^{\prime}_0$, $\gamma^{\prime}_1$, $\gamma^{\prime}_2$, $\gamma^{\prime}_3$ and $\kappa^{\prime}$ are the
bare parameters entering the $8\times 8$ Hamiltonian.
They are related to the effective mass of the conduction band ($m_c$) and the Luttinger parameters of the hole bands ($\gamma_{1,2,3}$ and $\kappa$) as
\begin{align}
\gamma^\prime_0 &= \gamma_0 - \frac{E_P}{E_g} \frac{E_g + \frac{2}{3}\Delta}{E_g + \Delta},\label{g0r}\\
\gamma^{\prime}_1 &= \gamma_1 - \frac{1}{3}\frac{E_P}{E_g},\label{g1r}\\
\gamma^{\prime}_2 &= \gamma_2 - \frac{1}{6}\frac{E_P}{E_g},\label{g2r}\\
\gamma^{\prime}_3 &= \gamma_3 - \frac{1}{6}\frac{E_P}{E_g}\label{g3r},\\
\kappa^{\prime} &= \kappa - \frac{1}{6}\frac{E_P}{E_g}\label{kr},
\end{align}
where
\begin{equation}
E_P = \frac{2m_0 P^2}{\hbar^2}\,, \quad \gamma_0 = \frac{m_0}{m_c}\,, \quad \gamma_0^\prime = \frac{m_0}{m_c^\prime}\,,
\end{equation}
and $E_g$ is a band gap.

These parameters are material specific and hence a function of the $z$-coordinate.
The order of operators in \eqref{eq:H8x8} is such that the Hamiltonian is indeed Hermitian.

The Hamiltonian~\eqref{eq:H8x8} exhibits unphysical solutions inside the band gap if $\gamma_0' < 0$.
These spurious solutions appear at large momenta, beyond the validity of the $\boldsymbol{k}\cdot\boldsymbol{p}$ model.
We apply the method put forward in Ref.~\cite{foreman_elimination_1997} to avoid these unphysical states: we renormalize $P$ in a way that $\gamma^\prime_0$ is equal to either $1$.
From~(\ref{g0r}) we thus obtain
\begin{equation}
 P^2 = \left(\gamma_0 - 1\right) \frac{E_g (E_g + \Delta)}{E_g + \frac{2}{3}\Delta} \frac{\hbar^2}{2m_0},
\end{equation}
which we then use to also renormalize the Luttinger parameters using~(\ref{g1r}-\ref{kr}).
With this renormalization, the spurious solutions at large $\boldsymbol{k}$ are pushed away from the band gap, whilst preserving the band structure around $\boldsymbol{k}=0$.

The band parameters for \InAsGaSb quantum wells are given in Table~\ref{table_pars_inasgasb}, we apply the renormalization method mentioned above to obtain the bare parameters of the $8 \times 8$ Kane Hamiltonian.
The valence band offsets are $0.56$ eV for GaSb-InAs, $0.18$ eV for AlSb-InAs, and $-0.38$ eV for AlSb-GaSb~\cite{halvorsen_optical_2000}.

The parameters for \HgTe quantum wells are given in Table~\ref{table_pars_hgte}.
Note that in this case already the bare parameters are given.
Those are such that they do not suffer from spurious solutions, and no renormalization procedure is necessary.

We calculate the parameters for the alloy $\textrm{Hg}_{0.3} \textrm{Cd}_{0.7} \textrm{Te}$ by linear interpolation of all Hamiltonian parameters except the band gap for which we use~\cite{pfeuffer-jeschke_transport_2000}
\begin{equation}
E_g (\textrm{eV}) = -0.303 (1-x) + 1.606 x - 0.132 x (1-x).
\end{equation}

The thickness of barrier materials, which we show in Fig.~\ref{fig:kp_disps} in the main text, are $5\,$nm for AlSb and $8\,$nm for (Hg,Cd)Te.

\begin{table*}
\caption{\label{table_pars_inasgasb}
Band structure parameters for InAs, GaSb, and AlSb~\cite{halvorsen_optical_2000, lawaetz_valence-band_1971}.
These parameters are the bare parameters and need to be renormalized before using them in simulation.
All parameters are for $T=0$ K.}
\begin{ruledtabular}
\begin{tabular}{c c c c c c c c c c}
& $E_g$ [eV] & $\Delta$ [eV]& $E_P$ [eV]& $m_c/m_0$ & $g_c$ & $\gamma_1$ & $\gamma_2$ & $\gamma_3$ & $\kappa$ \\
InAs & 0.41 & 0.38 & 22.2 & 0.024 & -14.8 & 19.67 & 8.37 & 9.29 & 7.68 \\
GaSb & 0.8128 & 0.752 & 22.4 & 0.042 & -7.12 & 11.80 & 4.03 & 5.26 & 3.18 \\
AlSb & 2.32 & 0.75 & 18.7 & 0.18 & 0.52 & 4.15 & 1.01 & 1.75 & 0.31 \\
\end{tabular}
\end{ruledtabular}
\end{table*}

\begin{table*}
\caption{\label{table_pars_hgte}
Band structure parameters for HgTe and CdTe~\cite{novik_band_2005, pfeuffer-jeschke_transport_2000}.
These parameters are already in renormalized form and can be used directly in the simulation.
Alloy parameters parameters for $\textrm{Hg}_{0.3} \textrm{Cd}_{0.7} \textrm{Te}$ are obtained using interpolation scheme from~\cite{pfeuffer-jeschke_transport_2000}.
All parameters are for $T=0$ K.}
\begin{ruledtabular}
\begin{tabular}{c c c c c c c c c c}
& $E_g$ [eV] & $\Delta$ [eV]& $E_P$ [eV]& $m_c^\prime/m_0$ & $g_c^\prime$ & $\gamma_1^\prime$ & $\gamma_2^\prime$ & $\gamma_3^\prime$ & $\kappa^\prime$ \\
HgTe & -0.303 & 1.08 & 18.8 & 1 & 2 & 4.1 & 0.5 & 1.3 & -0.4 \\
CdTe & 1.606 & 0.91 & 18.8 & 1.22 & 2 & 1.47 & -0.28 & 0.03 & -1.31 \\
$\textrm{Hg}_{0.3} \textrm{Cd}_{0.7} \textrm{Te}$ & 1.006 & 0.961 & 18.8 & 1.445 & $2$ & 2.259 & -0.046 & 0.411 & -1.037\\
\end{tabular}
\end{ruledtabular}
\end{table*}

We perform all \kp simulations by discretizing the Hamiltonian~\eqref{eq:H8x8} using a grid spacing of $a=0.5\,$nm.
For 2D bulk dispersion we only discretize $z$-direction, when calculating the edge dispersion we discretize both $y$- and $z$-directions.
We calculate all band structures by treating momentum $k_x$ as number, which we simple denote as $k$ in the figures.

In all simulations we consider magnetic field ${\boldsymbol{B} = B\hat{y}}$ along y-direction.
We include magnetic field through Zeeman and orbital effect.
\kp Zeeman term~\cite{Winkler} is
\begin{align}
H_{6c\,6c}^{z}&= \frac{1}{2} \, g^\prime \mu_{B}\,\boldsymbol{\sigma} \cdot \boldsymbol{B}\,,\\
H_{8v\,8v}^{z}&= -2 \, \mu_{B}\,\kappa^\prime \, \boldsymbol{J} \cdot \boldsymbol{B}\,,\nonumber\\
H_{7v\,7v}^{z}&= -2 \, \mu_{B}\,\kappa^\prime \, \boldsymbol{\sigma} \cdot \boldsymbol{B}\nonumber\,,\\
H_{8v\,7v}^{z}&= -3 \, \mu_{B}\,\kappa^\prime \, \boldsymbol{U} \cdot \boldsymbol{B}\nonumber\,,
\end{align}
where $\boldsymbol{\sigma}$ is a vector of Pauli matrices, $U = T^\dagger$, and
\begin{widetext}
\begin{equation}
T_x = \frac{1}{3\sqrt{2}}
  \begin{pmatrix}
    -\sqrt{3} & 0 & 1 & 0\\
    0 & -1 & 0 & \sqrt{3}\\
  \end{pmatrix}\,,\quad
T_y = \frac{-\textrm{i}}{3\sqrt{2}}
  \begin{pmatrix}
    \sqrt{3} & 0 & 1 & 0\\
    0 & 1 & 0 & \sqrt{3}\\
  \end{pmatrix}\,,\quad
T_z = \frac{\sqrt{2}}{3}
  \begin{pmatrix}
    0 & 1 & 0 & 0\\
    0 & 0 & 1 & 0\\
  \end{pmatrix}\,,
\end{equation}
\begin{equation}
J_x = \frac{1}{2}
\begin{pmatrix}
0 & \sqrt{3} & 0 & 0 \\
\sqrt{3} & 0 & 2 & 0 \\
0 & 2 & 0 & \sqrt{3} \\
0 & 0 & \sqrt{3} & 0\\
\end{pmatrix}\,,
\quad
J_y = \frac{\textrm{i}}{2}
\begin{pmatrix}
0 & -\sqrt{3} & 0 & 0\\
\sqrt{3} & 0 & -2 & 0\\
0 & 2 & 0 & -\sqrt{3}\\
0 & 0 & \sqrt{3} & 0\\
\end{pmatrix}\,,
\quad
J_z = \frac{\textrm{1}}{2}
\begin{pmatrix}
3 & 0 & 0 & 0\\
0 & 1 & 0 & 0\\
0 & 0 & -1 & 0\\
0 & 0 & 0 & -3\\
\end{pmatrix}\,.
\end{equation}
\end{widetext}
We add the orbital effect through a vector potential
\begin{equation}
A_x = B (z-z_0)\,,
\end{equation}
where $z_0$ is a coordinate offset which will be of relevance for finding effective models.
We include the vector potential in Hamiltonian by making the substitution
\begin{equation}
k_x \rightarrow k_x + \frac{2 \pi}{\phi_0} A_x\,,
\end{equation}
where $\phi_0=\frac{h}{e}$ is the flux quantum.
In the regime of parameters used in simulation this method gives the same results as using Peierl's substitution on the tight-binding level.
We decided to use for this route due to its advantages for obtaining effective models that we describe in next section.

\subsection{Effective low-energy models from \kp simulations}

We obtain effective models using quasi-degenerate perturbation theory, also known as L\"{o}wdin partitioning~\cite{lowdin_note_1951, luttinger_motion_1955, bir_symmetry_1974, Winkler}.
The main idea of this method is to choose group of states that we are interested in (group $A$) and treat all other states (group $B$) as perturbation.
In our case, $A$ will be the four (including spin) $k_x=k_y=0$ states closest to the topological gap at zero magnetic field (an electron-like and a heavy-hole state in both \InAsGaSb and \HgTe), giving an effective $4\times 4$ Hamiltonian.

We split our Hamiltonian into three parts
\begin{equation}
H = H_0 + H_1 + H_2,
\end{equation}
where $H_0$ is unperturbed Hamiltonian with known energies and wavefunctions, $H_1$ is perturbation that only acts between states from groups $A$ and $B$ separately, $H_2$ is perturbation that couples states from blocks $A$ and $B$.
We need to find unitary operator $e^{-S}$ that transform Hamiltonian into block-diagonal form
\begin{equation}
\label{eq: lowdin-transformation}
\tilde{H} = e^{-S} H \, e^{S}\,,
\end{equation}
with uncoupled blocks $A$ and $B$.

We base our implementation on equations (B.7) and (B.6) from~\cite{Winkler}:
\begin{widetext}
\begin{subequations}
\begin{eqnarray}
\tilde{H_d}
\label{eq: lowdin-hd}
    &=& \sum_{n=0}^{\infty}\frac{1}{(2n)!} [H^0 + H^1,\,S]^{(2n)}
    + \sum_{n=0}^{\infty}\frac{1}{(2n+1)!} [H^2,\,S]^{(2n+1)}\,,\\
\tilde{H_n}
\label{eq: lowdin-hn}
    &=& \sum_{n=0}^{\infty}\frac{1}{(2n+1)!} [H^0 + H^1,\,S]^{(2n+1)}
    + \sum_{n=0}^{\infty}\frac{1}{(2n)!} [H^2,\,S]^{(2n)}\,.
\end{eqnarray}
\end{subequations}
\end{widetext}
By requiring \eqref{eq: lowdin-hn} to vanish we find successive approximations of $S$ that we then use to solve \eqref{eq: lowdin-hd}.
We expand Hamiltonian $H$ into a polynomial which generators are parameters of our perturbation, e.g. momenta $k_x$ and $k_y$, and magnetic field $B_y$.
We truncate sum in \eqref{eq: lowdin-hd} to $n=3$ which allows us to obtain all terms up to $6th$ order in perturbation.
We then collect all terms that are polynomial up to desired order in perturbation parameters.
For example if we want to have effective model that is 2nd order on momenta we collect terms $k_x$, $k_y$, $k_x^2$, $k_y^2$, and $k_xk_y$.
This gives us the effective model that describes dispersion of exact model up to desired precision.

\section{Higher-order effective models for \InAsGaSb}

\begin{figure}
\includegraphics[width=\columnwidth]{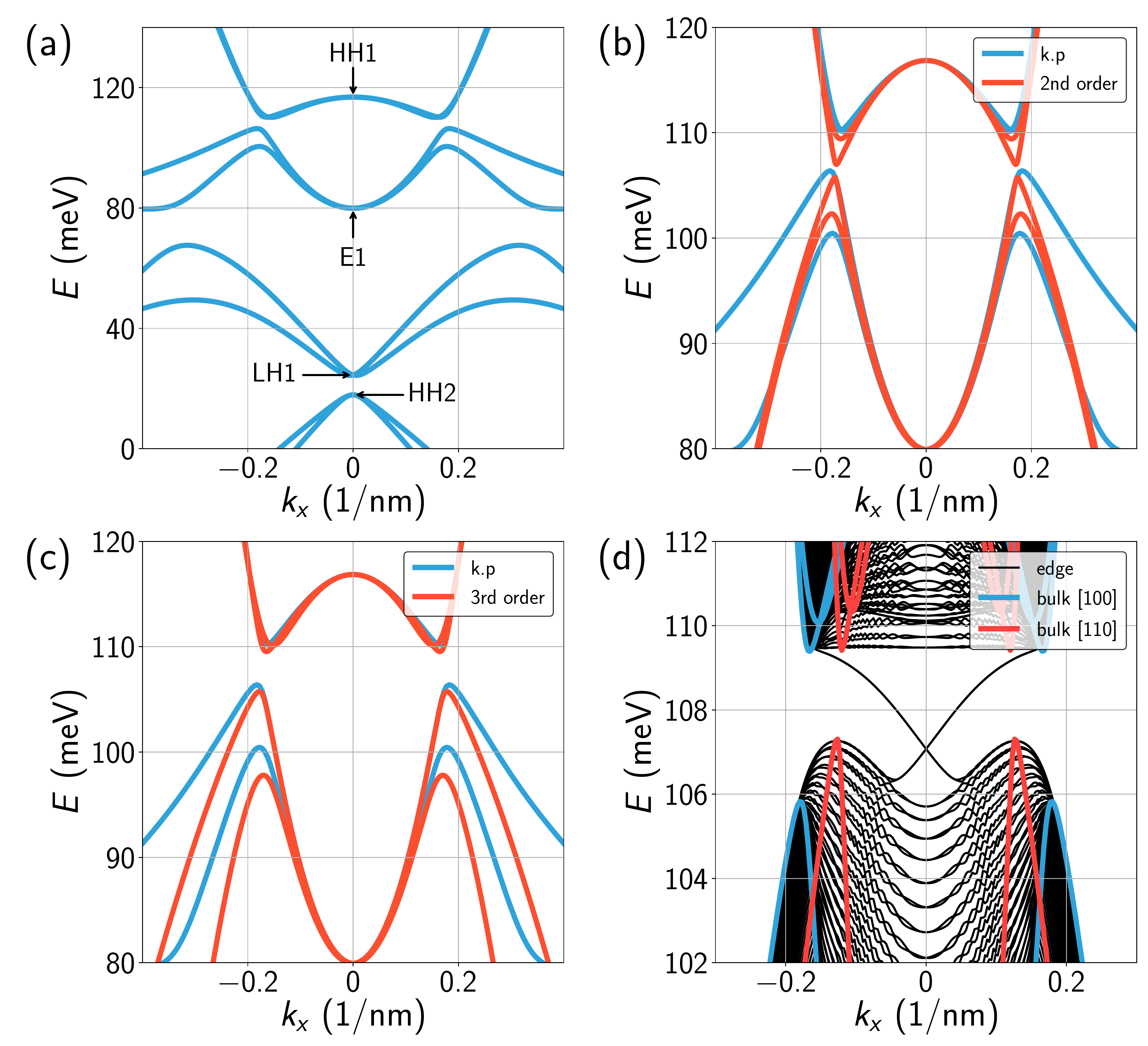}
\caption{
Effective models for InAs/GaSb quantum well with layer thickness $12.5\,$nm/$5\,$nm.
(a) \kp dispersion of 2D system with
labelled bands.
We compare 2nd order (b) and 3rd (c) continuous dispersions of effective models with exact \kp dispersion from plot (a).
We see that 2nd order effective model underestimates the topological gap, and therefore does not describe properly investigated system.
(d) Tight-binding dispersion of bulk (blue) and edge (black) system made from 3rd order effective model.
We observe that Dirac point is buried in the valence band.
We discretized effective model with grid spacing $a=2\,$nm.
Width of the system used to simulate edge states is $W=800\,$nm.
}
\label{fig:effective-models}
\end{figure}

Using the L\"owdin partitioning technique detailed above, we have derived effective models for \InAsGaSb quantum well with layer thickness {$12.5\,$nm/$5\,$nm} on Fig.~\ref{fig:effective-models}.

In Fig.~\ref{fig:effective-models}(a) we show the bulk \kp band structure of this quantum well on a larger energy range.
As for the BHZ model, we choose the electron-like state E1 and the heavy-hole state HH1 as the basis of our perturbation theory.
Other hole states such as LH1 and HH2 are close, but still further away in energy than the inversion gap.
Still, as we will see, they have a significant influence.

We have numerically derived $4\times 4$ effective models with momenta up to second order (this is equivalent to the BHZ model including linear and quadratic spin-orbit terms similar to~\cite{Rothe2010}) and third order.
The comparison of the full \kp band structure with the second order model in Fig.~\ref{fig:effective-models}(b) shows the limits of this approximation clearly: in particular, the hybridization gap is far too small.
Only after including third order terms (Fig.~\ref{fig:effective-models}(c)) do we find a satisfactory agreement.
These third-order terms are (at least partially) due to interactions with hole states that are further away in energy.
These still have a significant influence on the band structure at finite momentum.

In ~Fig.~\ref{fig:effective-models}(d) we show the dispersion in a strip of finite width $W$ (black lines), with edges along the [100] direction.
We observe that the Dirac point of the edge states is clearly buried in the bulk valence band.
In particular, we observe that this burying is due to the anisotropy of the hole band structure: the hybridization gap in [110] direction is very different from the hybridization gap in [100] direction.
As seen above, to describe this anisotropy faithfully, we needed to take into account the further-away hole bands in the form of higher-order momentum terms.

\subsection{Derivation of the suppression of the edge-state $g$-factor}

To derive the edge-state wavefunction we start from the BHZ model \eqref{eq:BHZ1}, \eqref{eq:BHZ2}.
In this section we re-derive the results of~\cite{Zhou2008}, on which we build our $g$-factor derivation.
The wavefunctions for the edge states decay into the bulk and can be written as
\begin{align}
\Psi_{1, 2} = \Psi_ e^{\pm \lambda_{1,2} y},
\end{align}
where:
\begin{align}
\lambda_{1,2}^2 = k_x + F \pm \sqrt{F^2 - (M^2 - E^2)/B_+ B_-}, \\
F = \frac{A^2 - 2(MB + ED)}{2 B_+ B_-}.
\end{align}
Plugging this back into the BHZ model \eqref{eq:BHZ1} and \eqref{eq:BHZ2}:
\begin{align}
[M - B_+ (k_x^2 - \lambda_{1,2}^2)]\psi_1 + A (k_x \mp \lambda_{1,2})\psi_2 = E \psi_1;\\
A (k_x \pm \lambda_{1,2})\psi_1 - [M - B_- (k_x^2 - \lambda_{1,2}^2)]\psi_2 = E \psi_2.
\end{align}
The two spin sectors are different by the sign of $A$, therefore the decay length is the same for the opposite spins.

Let us now solve for the decaying solutions in half-space $y>0$ with hard-wall boundary conditions at $y=0$.
The condition that the wavefunction can vanish at the hard wall is the same as requiring linear dependence of the decaying solutions at $y=0$:
\begin{align}
(\psi_1/\psi_2)_{1} = (\psi_1/\psi_2)_{2},
\end{align}
where outer index is enumerating the decaying solutions  $\Psi_{1, 2}$.
Therefore:
\begin{align}
\frac{M + E - B_-(k_x^2 - \lambda_1^2)}{A (k_x - \lambda_1)} =
\frac{M + E - B_-(k_x^2 - \lambda_2^2)}{A (k_x - \lambda_2)}.
\label{eq:ratios1}
\end{align}
Let us solve the equation for the crossing point of the edge dispersion, where due to time-reversal symmetry $k_x=0$, therefore:
\begin{align}
\lambda_2(M + E + B_0 \lambda_1^2) =
\lambda_1(M + E + B_0 \lambda_2^2).
\end{align}
Then we use that $\lambda_1 \lambda_2 = \sqrt{\frac{M^2 - E^2}{B_+ B_-}}$, and get:
\begin{align}
(E + M) \lambda_2 + B_- \lambda_1 \sqrt{\frac{M^2 - E^2}{B_+ B_-}} \nonumber \\  =(E + M) \lambda_1 + B_- \lambda_2 \sqrt{\frac{M^2 - E^2}{B_+ B_-}}.
\end{align}
This equation has a solution if:
\begin{align}
E + M - \sqrt{B_-/B_+} \sqrt{M^2 - E^2} = 0.\label{eq:E_condition}
\end{align}
Therefore, the crossing is at $E=- M\frac{D}{B}$.
Note that the result has correct limit $E=0$ when the bandstructure is symmetric, $D=0$.

We now proceed with the solution by computing the matrix element of the Zeeman energy between the two edge states at the crossing point, where the gap is opened.
Let us denote $\psi_2/\psi_1 = r$.
For opposite spin the ratio is $\psi_2'/\psi_1' = r^*$, therefore if we use the $g_e$ and $g_h$ (electron and hole bulk in-plane $g$-factors), the effective edge $g$-factor is:
\begin{align}
g_{eff} = \frac{g_e + g_h |r|^2}{1 + |r|^2}.
\end{align}
For the parameters of the crossing point ($E = - MD/B$, $k_x=0$) we get from \eqref{eq:ratios1} and \eqref{eq:E_condition}:
\begin{align}
|r|^2 = \left|\frac{M B_+ + B B_+ \lambda_1^2}{M B_- + B B_- \lambda_1^2}\right| = \frac{B_+}{B_-}.
\end{align}
This gives simply \eqref{eq: edge-g-factor} from the main text.

\subsection{Numerical values for $g$-factors in \InAsGaSb and
\HgTeCdTe quantum wells}

Using L\"owdin partitioning, we compute the $g$-factor of the electron states in the $4 \times 4$ model, by doing perturbation in $B_y$ instead of momenta.
This is a gauge-invariant quantity in the \HgTe quantum wells due to inversion symmetry.
However, it becomes gauge-dependent in \InAsGaSb due to the linear spin-orbit terms that are essential in this strongly asymmetric structure.
We fix the gauge $z_0$ by demanding that the off-diagonal matrix elements between E1 and HH1 do not depend on $B$.
This is the same gauge as used in~\cite{Hu2016a}.

\begin{figure}
\includegraphics[width=\columnwidth]{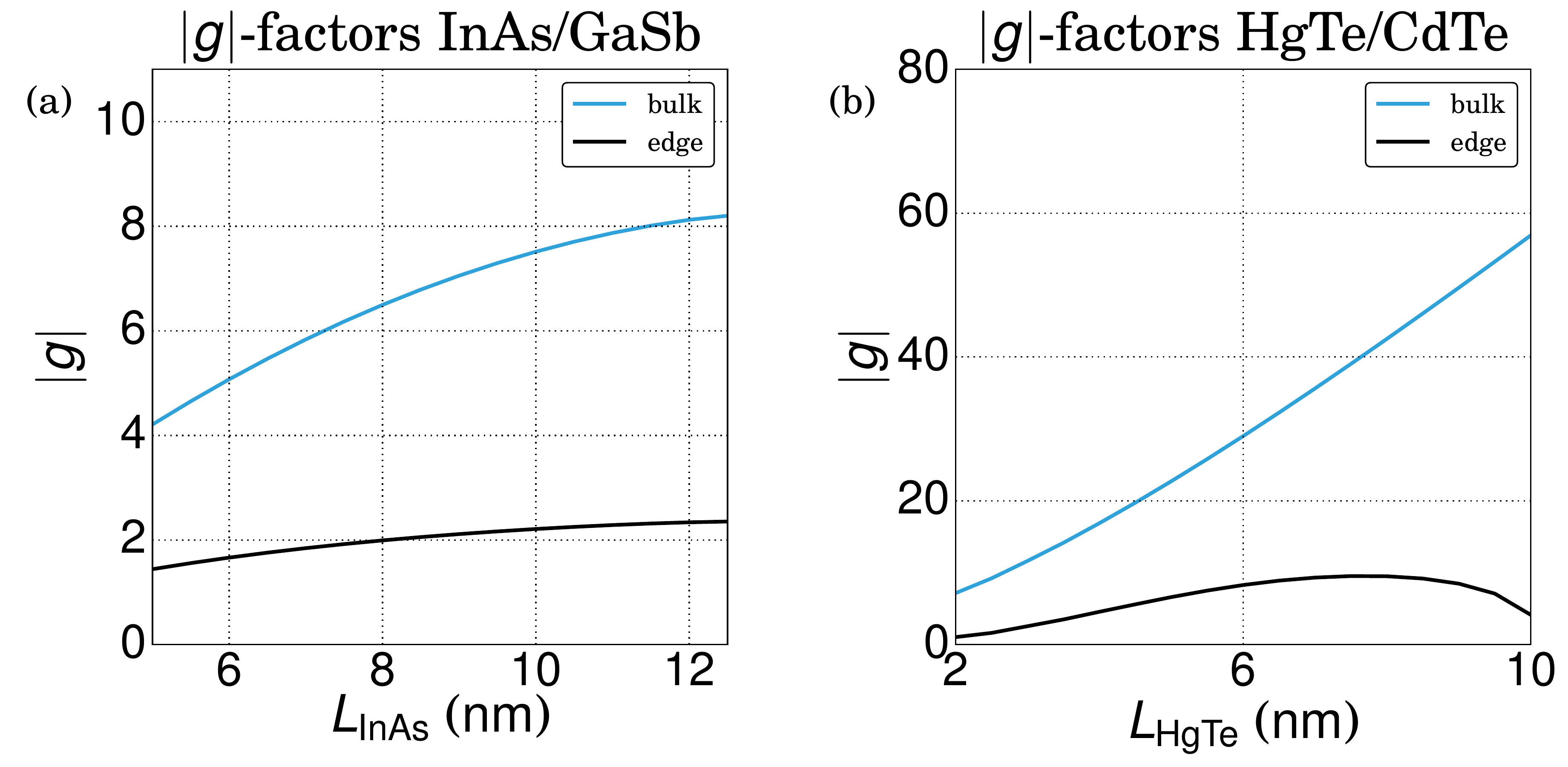}
\caption{
bulk and effective edge-state $g$-factors for (a) $\mathrm{InAs}/\mathrm{GaSb}$ and (b) $\mathrm{HgTe}/\mathrm{CdTe}$ quantum wells.
}
\label{fig:g-factors}
\end{figure}

In Fig.~\ref{fig:g-factors} we show the bulk $g$-factors as a function of quantum well width for both types of quantum wells, together with the value of the effective edge $g$-factor obtained from \eqref{eq: edge-g-factor} using parameters $B$ and $D$ from the derived effective model.
As discussed in the main text we observe a strong suppression of the effective edge $g$-factor compared to the bulk value.

\end{document}